\documentclass[12pt]{article}
\usepackage[dvips]{graphics}

\def\PR#1#2#3{Phys. Rev. {\bf #1}, #2 (#3)}
\def\PRL#1#2#3{Phys. Rev. Lett. {\bf #1}, #2 (#3)}
\def\PL#1#2#3{Phys. Lett. {\bf #1}, #2 (#3)}

\def\NP#1#2#3{Nucl. Phys. {\bf #1}, #2 (#3)}

\def\PTP#1#2#3{Prog. Theor. Phys. {\bf #1}, #2 (#3)}
\def\EPJ#1#2#3{Eur. Phys. J. {\bf #1}, #2 (#3)}

\def\mz{m_Z^{}}
\def\mr{m_R^{}}
\def\Su{\rm u}
\def\Sd{\rm d}

\def\he{{y^{\rm e}}}

\def\nn{\nonumber}
\def\dfrac{\displaystyle\frac}
\def\eqref#1{eq.(\ref{eqn:#1})}

\def\eqlab#1{\label{eqn:#1}}
\def\vev#1{\langle #1 \rangle}

\begin{document}

\title{
Analytic Solutions to the RG Equations\\
of the Neutrino Physical Parameters
}
\author{
{N. Haba$^{1}$}\thanks{E-mail address:haba@eken.phys.nagoya-u.ac.jp}
{, Y. Matsui$^2$}\thanks{E-mail address:matsui@eken.phys.nagoya-u.ac.jp}
{ and N. Okamura$^3$}\thanks{E-mail address:naotoshi.okamura@kek.jp}
\\
\\
\\
{\small \it $^1$Faculty of Engineering, Mie University,}
{\small \it Tsu Mie 514-8507, Japan}\\
{\small \it $^2$Department of Physics, Nagoya University,}
{\small \it Nagoya, 464-8602, Japan}\\
{\small \it $^3$Theory Group, KEK, Tsukuba Ibaraki 305-0801, Japan}}
\date{}
\maketitle


\vspace{-10.5cm}
\begin{flushright}
hep-ph/9911481\\
DPNU-99-31\\
KEK-TH-662
\end{flushright}
\vspace{10.5cm}
\vspace{-2.5cm}

%
%

\begin{abstract}

In the case of two generation neutrinos,
 the energy-scale dependence of the
 lepton-flavor mixing matrix with Majorana phase can be governed
 by only one parameter $r$,
 which is the ratio between the diagonal elements of neutrino mass matrix.
By using this parameter $r$,
we derive the analytic solutions
to the renormalization group equations of
the physical parameters,
which are the mixing angle, Majorana phase,
and the ratio of the mass-squared difference
to the mass squared of the heaviest neutrino.
 The energy-scale dependence of the Majorana phase is clarified
by using these analytic solutions.
 The instability of the Majorana phase causes in the same parameter
region in which the mixing angle is unstable against quantum corrections.

\end{abstract}

\newpage

%
%

\section{Introduction}

Recent neutrino oscillation experiments suggest
 the strong evidences of the tiny neutrino masses
 and the lepton flavor mixing \cite{solar4}-\cite{LSND}.
Studies of the lepton flavor mixing matrix,
 which we call Maki-Nakagawa-Sakata (MNS) matrix \cite{MNS},
 will give us important cues of the physics
 beyond the standard model.
One of the most important studies of the
 lepton-flavor mixing is
 the analysis of quantum corrections of
 the MNS matrix \cite{up2now}-\cite{EL1}.
Especially,
 the Majorana phase plays the crucial role
 for stabilities of the mixing angle
 against quantum corrections \cite{HMOS2}.

In this paper we analyze
 the energy-scale dependence of the MNS matrix
 with the physical Majorana phase
 in two generation neutrinos.
According to the LSND experiment\cite{LSND},
the scenario of two heavy degenerate neutrinos
 can be realistic,
 where we can neglect
 the first generation effects in the
 energy-scale dependence of the MNS matrix.
In this scenario,
 the MNS matrix has two parameters, which are
 the mixing angle and the Majorana phase.
The energy-scale dependence of them
 are completely determined
 by only one parameter $r$
 which is the ratio between the diagonal elements
 of neutrino mass matrix.
The energy-scale dependence of $r$ can
 be easily solved.
When we use this parameter $r$,
 we can easily obtain the analytic solution
 to the energy-scale dependence
 of physical parameters in the MNS matrix
 which are the mixing angle and the Majorana phase.
The energy-scale dependence of $\Delta m^2_{23} / m_3^2$
 is also determined by only one parameter $r$,
 where $\Delta m^2_{23}$ is the mass-squared difference
 between the second and the third generations
 and $m_3$ is the absolute value of the heaviest
 neutrino mass.
 The energy-scale dependence of the Majorana phase is clarified
by using these analytic solutions.
 The instability of the Majorana phase causes in the same parameter
region in which the mixing angle is unstable against quantum corrections.

%
%

\section{Energy-Scale Dependence of $\kappa$}

 The effective Yukawa couplings in the lepton sector are given by
\begin{equation}
{\cal
L}^{\rm low}_{yukawa}
           =  {\he}_{ij} \phi_{\Sd} L_i \cdot e_{Rj}^c
            - \dfrac{1}{2} \kappa_{ij}(\phi_{\Su}L_i)\cdot(\phi_{\Su}L_j)
            + h.c.\,,
\eqlab{W_low}
\end{equation}
where
$\phi_{\Su}$ and $\phi_{\Sd}$
are the SU(2)$_L$ doublet Higgs bosons
that give Dirac masses to the up-type and down-type
fermions, respectively.
 $L_i$ is the $i$-th generation SU(2)$_L$
doublet lepton.
 $e_{Ri}^{}$ is the $i$-th generation right-handed charged-lepton.
 The matrix $\kappa$ induces the neutrino Majorana mass matrix.
${y^{\rm e}_{}}$ is the Yukawa matrix of the charged-lepton.
 In this paper, we take the diagonal base of ${y^{\rm e}_{}}$.
 Once ${y^{\rm e}_{}}$ is taken diagonal at a certain scale,
the diagonality of it is kept at all energies in the one-loop level.
This is because there are no lepton-flavor-mixing terms, except for
$\kappa$ in the minimal supersymmetric standard model (MSSM).
 In this base the matrix $\kappa$ is
diagonalized as
\begin{equation}
   \kappa_{D}^{} = U^T \kappa \; U \,,
\end{equation}
where $U$ is an unitary matrix (the MNS matrix)
and $\kappa_{D}^{}$ is the diagonalized mass matrix.
The MNS matrix $U$ is parameterized as,
\begin{equation}
   U = e^{i x}
   \left(
   \begin{array}{cc}
   1  & 0 \\
   0  & e^{iy}
   \end{array}
   \right)
   \left(
   \begin{array}{cc}
    \cos \theta_{23} & \sin \theta_{23} \\
   -\sin \theta_{23} & \cos \theta_{23}
   \end{array}
   \right)
   \left(
   \begin{array}{cc}
   1  & 0 \\
   0  & e^{i \phi/2}
   \end{array}
   \right) \,,
\end{equation}
where $\theta_{23}$ is the mixing angle and $\phi$ is
the Majorana phase. The phases of $x$ and $y$ are unphysical parameters, which
 can be rotated out by the field redefinition. 

Since $\kappa$ is complex and symmetric matrix,
$\kappa$ can be parameterized as \cite{HMOS1}
\begin{equation}
   \kappa = \lambda
   \left( 
   \begin{array}{cc}
   r                    & c \sqrt{r}e^{i \chi} \\
   c \sqrt{r}e^{i \chi} & 1
   \end{array}
   \right)\, .
   \label{eq:param1}
\end{equation}
In this parameterization,
$c$ and $\chi$ are non-negative values 
and energy-scale independent parameters.
The energy-scale dependence of the $\kappa$ can be
controlled by two parameters $r$ and $\lambda$.
Since the $\lambda$ is overall factor,
the energy-scale dependence of the mixing angle and
the Majorana phase
in the MNS matrix are governed by only one parameter $r$.

The renormalization group equation for $r$ is given as \cite{HMOS1,HO1}
\begin{equation}
   \dfrac{d}{dt} \ln r = - \dfrac{1}{8 \pi^2}(y_{\tau}^2 - y_{\mu}^2),
   \label{RGEr1}
\end{equation}
where $t$ is the scaling parameter which is related to the 
renormalization scale $\mu$ as $ t = \ln \mu$.
$y_{\tau}$ and $y_{\mu}$ are Yukawa couplings of the charged-leptons,
$\tau$ and $\mu$, respectively. 
We can obtain the solution to eq.(\ref{RGEr1}) as
\begin{equation}
r(\mr)=r(\mz)
\left(1- \epsilon
\right)^2\,,
\end{equation}
where $\mr$ is a certain high energy scale.
When $\tan \beta \leq 50$,
where $\tan \beta$ is the ratio of two vacuum expectation
values of Higgs bosons 
$\tan \beta = \vev{\phi_{\Su}}/\vev{\phi_{\Sd}}$,
the parameter $\epsilon$ can be obtained
approximately \cite{HO1} as
\begin{eqnarray}
\epsilon &\simeq& 1-\exp
\left(-
\dfrac{1}{16\pi^2}
\int_{\ln\mz}^{\ln \mr}
\left(y_\tau^2-y_\mu^2\right)
dt
\right)\,, \nn \\
\nn \\
&\simeq&
\dfrac{y_{\tau}^2}{16 \pi^2}
                   \ln \left( \dfrac{\mr}{\mz} \right)\,.
   \label{RGEe1}
\end{eqnarray}
\begin{figure}[t]
   \begin{center}
   \scalebox{0.8}{\includegraphics{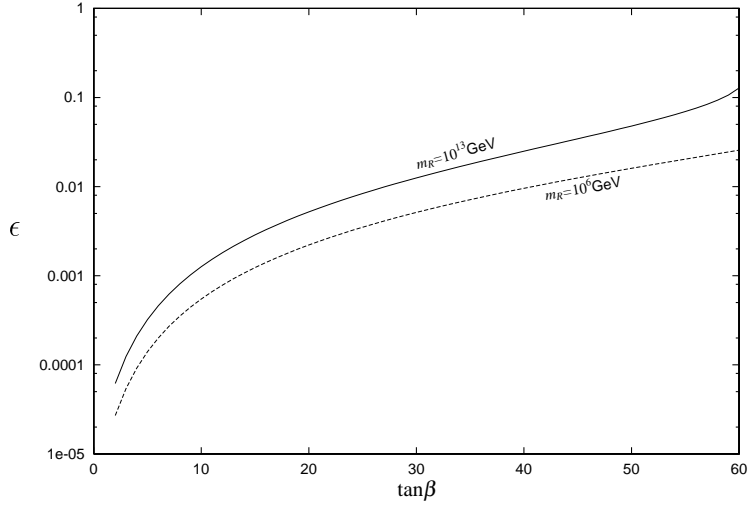}}
   \end{center}
   \caption{$\tan \beta$ dependence of the $\epsilon$ at
 $\mr=10^6$GeV and $\mr=10^{13}$GeV. }
   \label{Fig:epsilon}
\end{figure}
Figure~{\ref{Fig:epsilon}} shows
the $\tan \beta$ dependence of $\epsilon$
at $\mr=10^6$GeV and $\mr=10^{13}$GeV.
Since the energy-scale dependence of $r$ is very small,
the difference between
the values of $r(\mr)$ and $r(\mz)$ is not so large
\cite{HMOS1,HO1}.
For instance, in the case of
$\tan \beta = 20$ and $\mr = 10^{13} \; \mbox{GeV}$,
$\epsilon$ is nearly equal to $0.005$.
\vspace{24pt}

%
%

\section{Analytic Solutions to the RG Equations}
By using the energy-scale dependent parameters $r$ and $\lambda$, 
 and the energy-scale independent parameters $c$ and $\chi$,
 the masses of neutrinos, $\nu_2^{}$ and $\nu_3^{}$,
 are given as
\begin{eqnarray}
   m_2 & = & \dfrac{\lambda v^2 \sin^2 \beta}{4} 
   \sqrt{2 \left\{ r^2 
   + 1 + 2 c^2 r - f(r,c,\chi) \right\}} \,,
   \nonumber \\
   \\
   m_3 & = & \dfrac{\lambda v^2 \sin^2 \beta}{4}
   \sqrt{2 \left\{ r^2 + 1 + 2 c^2 r + f(r,c,\chi) \right\} } \,.
   \nonumber
\end{eqnarray}
The function $f(r,c,\chi)$ is defined as
\begin{equation}
   f(r,c,\chi)  \equiv  \sqrt{ (1 - r^2)^2 + 4 c^2 r \, \{g(r,\chi)\}^2}
   \,\,,
\end{equation}
where the function $g(r,\chi)$ is defined as
\begin{eqnarray}
   g(r,\chi) & \equiv & \left| \, r e^{- i \chi} + e^{i \chi} \,\right|
   \nonumber \\
   &=& \sqrt{ (r - 1)^2  + 2 r ( \cos 2 \chi + 1)} \,.
\end{eqnarray}
The mass-squared difference between neutrinos is given as
\begin{equation}
   {\Delta} m^2_{23} = m_3^2 - m_2^2
  = \lambda^2 f(r,c,\chi) \dfrac{v^4 \sin^4 \beta}{4}\,\,.
\end{equation}
The mass-squared difference scaled by
the heaviest neutrino mass $m_3^{}$
can be given as
\begin{equation}
   \dfrac{{\Delta} m_{23}^2}{m_3^2} =
   \dfrac{2 f(r,c,\chi)}{(r-1)^2 + 2 r ( c^2 + 1) + f(r,c,\chi)} \,\,.
   \label{deg1}
\end{equation}

By using the parameters, $c$, $\chi$, and $r$,
the mixing angle $\theta_{23}$ and the
Majorana phase $\phi$ can be written as
\begin{eqnarray}
   \sin^2 2 \theta_{23}
   & = &
     \dfrac{4 c^2 r \{g(r,\chi)\}^2}{\{f(r,c,\chi)\}^2}\,, \nonumber \\
   & = & \dfrac{1}
      {1 + {\displaystyle \dfrac{(r - 1)^2 (r + 1)^2}
         {4 c^2 r \{ (r - 1)^2 + 2 r (1 + \cos 2 \chi)\} }}
      }\,\,,
   \label{mix1}
\end{eqnarray}
and
\begin{equation}
   \cos \phi =
   \dfrac{{\displaystyle \cos 2\chi - c^2
   + \dfrac{2 r}{\{g(r,\chi)\}^2} \sin^2 2\chi }}
   {\sqrt{c^4 - 2 c^2 \cos 2\chi + 1}}\,\, .
   \label{Majo1}
\end{equation}
\vspace{24pt}

%
%

\section{Energy-Scale Dependence of the Physical Parameters}
The energy-scale dependent parameter is only $r$
in the physical parameters of
eqs.(\ref{deg1}), (\ref{mix1}) and (\ref{Majo1}).
Thus studying the energy-scale dependence of the physical parameters
is equivalent to studying the $r$ dependence of them.

At the low energy-scale, the large mixing between the
second and the third generations, $\sin^2 2 \theta_{23} \simeq 1$
is favored by the atmospheric neutrino experiments \cite{SK4}.
This means that
\begin{equation}
    \dfrac{(r - 1)^2 (r + 1)^2}
    {4 c^2 r \{ ( r - 1)^2 + 2 r ( 1 + \cos 2 \chi) \}} \ll 1 \,\,
    \label{eq:cond1}
\end{equation}
should be required at the low energy-scale from the eq.(\ref{mix1}).
There are following two cases which satisfy
the eq.(\ref{eq:cond1}):
\vspace{12pt}

\noindent
(a) The parameter $c$ is very large such as
\begin{equation}
    c^2 \gg  \dfrac{(r - 1)^2 (r + 1)^2}
    {4 r \{ ( r - 1)^2 + 2 r ( 1 + \cos 2 \chi) \}} \,\,.
    \label{eq:cond2}
\end{equation}
In this case,
the energy-scale dependence of the physical parameters
is negligible,
because the dominant component of the $\kappa$
is the energy-scale independent parameter $c$.
However,
even if the parameter $c$ is enough large but finite, 
eq.(\ref{eq:cond2}) is not satisfied
in the vicinity of $r=0$.
In this parameter region, the large mixing is unstable
against quantum corrections.
When $r$ is in the vicinity of the zero,
the parameter $c$ must be infinitely large in 
order to satisfy eq.(\ref{eq:cond2}).
Equation(\ref{eq:cond2}) can be rewritten as
\begin{equation}
 c\sqrt{r} \gg \dfrac{(r - 1)(r + 1)}
{2g(r,\chi)} \,,
\eqlab{casea1}
\end{equation}
where $c\sqrt{r}$ is the absolute value of the off-diagonal
elements in the matrix $\kappa$.
When \eqref{casea1} is satisfied,
 the large mixing is always preserved against quantum corrections
\cite{HO1,EL1}.
This case corresponds with so-called pseudo-Dirac type neutrino 
 mass matrix \cite{PD}.
\vspace{12pt}

\noindent
(b) The parameter $r$ is nearly equal to 1
with $\chi\neq\pi/2$.
We discuss the case of $\chi=\pi/2$ later.
In this case,
it is necessary that we study the $r$ dependence of the
physical parameters
in the vicinity of the $r=1$,
because mixing angle $\sin^2 2 \theta_{23}^{}$  
is not stable against
quantum corrections in this region \cite{HOS1}.
Since the large $c$ makes the mixing angle 
to be stable against quantum corrections
as we have seen in the case (a),
we study the region of $c<10$.
\begin{figure}[t]
   \begin{center}
   \scalebox{0.8}{\includegraphics{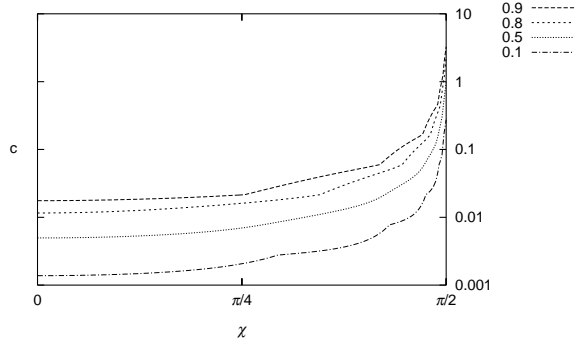}}
   \end{center}
   \caption{The contour plot of $\sin^2 \theta_{23}$ at $\epsilon = 0.005$
            ($\mr=10^{13}$GeV and $\tan \beta =20$). }
   \label{Fig:Mixing1}
\end{figure}
Here we input $\sin^2 2\theta_{23}^{}=1$ at $\mz$ scale.
Figure \ref{Fig:Mixing1} shows the contour plot of
$\sin^2 2 \theta_{23}$ at $\mr$ scale ($\epsilon = 0.005$)
for the continuous changes of $c$ and $\chi$.
The more the value deviates from 1
the larger the change of the mixing angle 
by quantum corrections
becomes.
 As the values of $\chi$ approach $\pi/2$ and the parameter $c$ 
becomes small, the $\sin^2 2 \theta_{23}$ becomes unstable
against quantum correction.
 When we define $\alpha\equiv\pi/2 - \chi$ and take
the parameters, $c$, $\epsilon$ and $\alpha$ to be small,
the left-hand side of eq.(\ref{eq:cond1}) can be rewritten as
\begin{equation}
    \dfrac{(r - 1)^2 (r + 1)^2}
    {4 c^2 r \{ ( r - 1)^2 + 2 r ( 1 + \cos 2 \chi) \}} 
    \sim \dfrac{\epsilon^2}{c^2 \alpha^2 } \,\,.
\eqlab{appcaseb}
\end{equation}
 This indicates that 
if the $c \alpha$ is much smaller than $\epsilon$,
eq.(\ref{eq:cond1}) is not satisfied.
 Therefor the mixing angle is unstable
against the quantum corrections in this region.
This result is the same as that of Ref.~\cite{HO1,HOS1,EL1}.

In order to understand
the physical meaning of the parameters $r$($\epsilon$), 
$c$ and $\chi$($\alpha$) obviously,
these parameters should 
be related to the other physical parameters, 
the Majorana phase and the mass-squared difference.
 Especially, it is necessary to investigate 
the physical meaning of the parameter 
regions in which the mixing angle $\sin^2 2 \theta_{23}$
is unstable against quantum corrections.

\begin{figure}[t]
   \begin{center}
   \scalebox{0.5}{\includegraphics{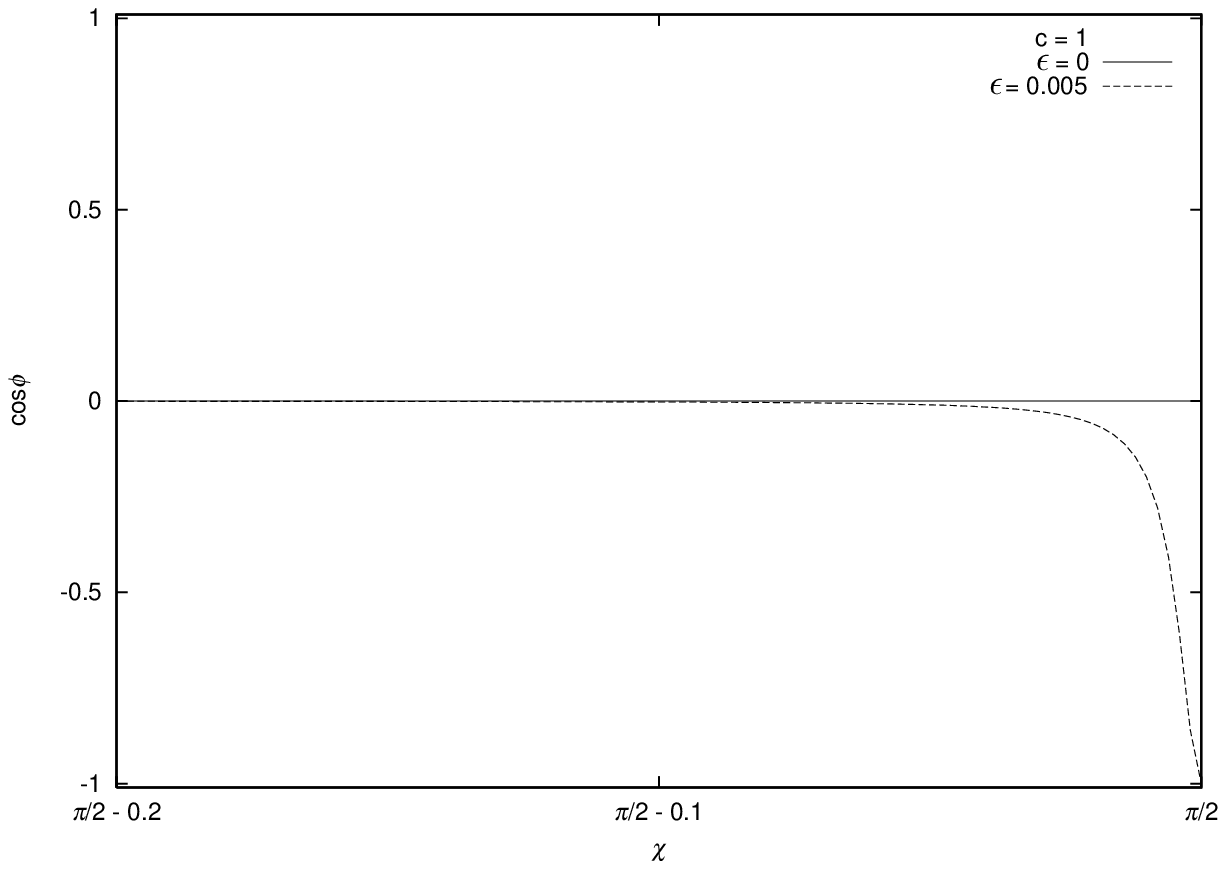}}
   \scalebox{0.5}{\includegraphics{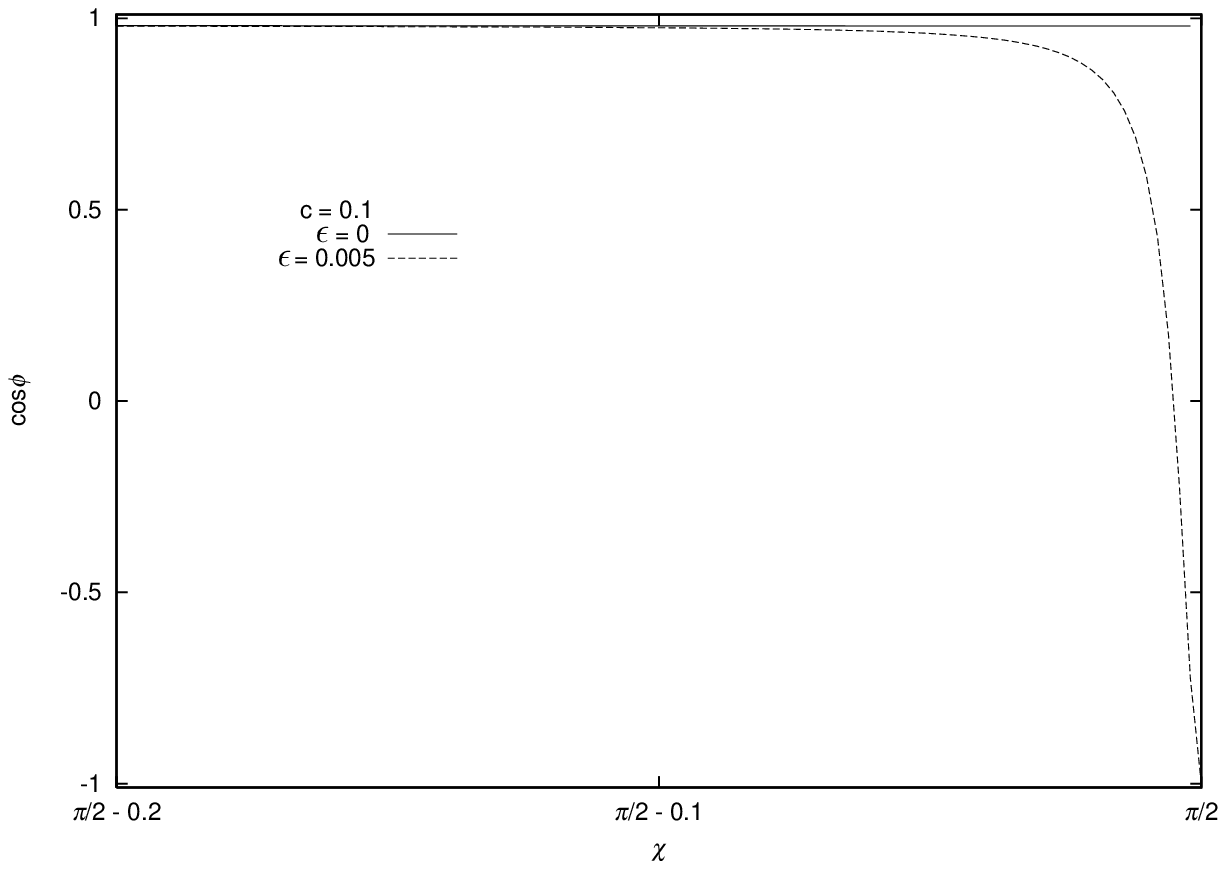}}
   \end{center}
   \caption{The plot of the Majorana phase $\phi$ at
   $\epsilon = 0$ and $\epsilon = 0.005$.}
   \label{Fig:Majorana1}
\end{figure}

Figure \ref{Fig:Majorana1} shows 
the Majorana phase $\phi$ in eq.(\ref{Majo1}) 
in the vicinity of $\chi = \pi/2$ 
for $c = 1$ and $c = 0.1$ cases,
when we take $\epsilon = 0$ and $\epsilon = 0.005$ for both cases.
We set 
the $\sin^2 2 \theta_{23} =1$ at the $\mz$ scale.
 Figure \ref{Fig:Majorana1} shows that
the Majorana phase is not changed by the quantum corrections,
when the $\chi$ is not close to $\pi/2$.
 By using small parameter $\alpha$,
the Majorana phase is obtained as
\begin{equation}
\cos \phi = \dfrac{1-c^2}{1+c^2}
\left(1+\dfrac{2c^2}{\left(c^2+1\right)^2}\alpha^2\right)
\end{equation}
for $r=1$.
By using this equation, we obtain the Majorana phase
at the $\mz$ scale for the large mixing. 
On the other hand,
 when the $\chi$ is close to the $\pi/2$,
 the Majorana phase is changed by the quantum corrections
 and it always becomes close to $-1$. 
The mixing angle is unstable against
quantum corrections
in the region of small $c$ and
$\chi \simeq \pi/2$.
Comparing to Figure \ref{Fig:Mixing1}
and Figure \ref{Fig:Majorana1},
we notice that
the instability of the Majorana phase causes 
in the same parameter region in which 
the mixing angle is unstable against quantum corrections.

\begin{figure}[ht]
   \begin{center}
   \scalebox{0.75}{\includegraphics{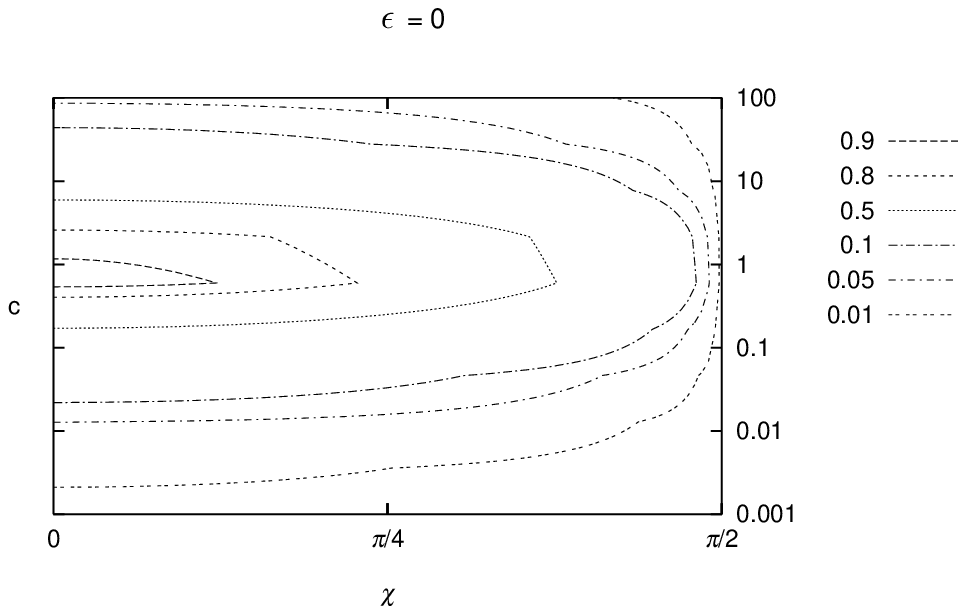}}
   \scalebox{0.75}{\includegraphics{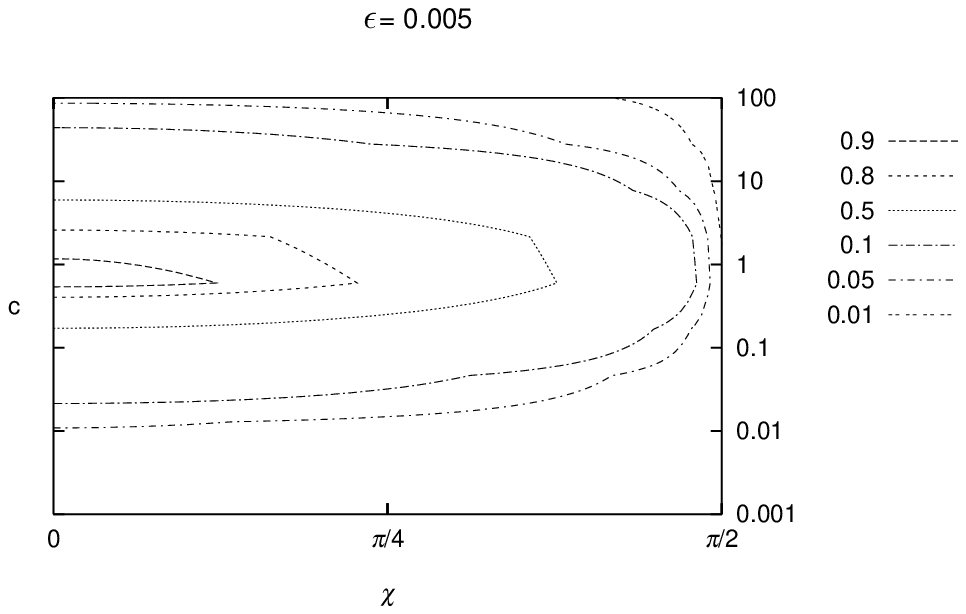}}
   \end{center}
   \caption{The contour plots of $\Delta m_{23}^2/m_3^2$ at
   $\epsilon = 1$ and $\epsilon = 0.005$.}
   \label{Fig:Massdiff1}
\end{figure}

Figure \ref{Fig:Massdiff1} shows the contour plot of
${\Delta} m_{23}^2 / m_3^2$ in eq.(\ref{deg1}) 
at $\epsilon = 0$ and 
$\epsilon = 0.005$ for the continuous
changes of $c$ and $\chi$.
According to the Figure \ref{Fig:Massdiff1},
the parameter region
in which the mixing angle is unstable against
quantum corrections is corresponding to
the region in which the masses are more degenerate.
This result is consistent with that of Ref.~\cite{HOS1}.
\vspace{24pt}

 Finally, we discuss the special case of $\chi=\pi/2$,
where the $r$ dependence of the mixing angle
and the Majorana phase $\phi$ are quite different
from the case of $\chi\neq\pi/2$.
 In this case
eq.(\ref{mix1}) becomes
\begin{equation}
    \sin^2 2 \theta_{23} = \dfrac{1}
    {1 + {\displaystyle \dfrac{(r + 1)^2}
         {4 c^2 r}}
    }\,.
    \label{eq:mix3}
\end{equation}
 This equation indicates that $\sin^2 2 \theta_{23}$ 
does not automatically take the maximal mixing at $r = 1$.
It is because the mixing angle becomes 
\begin{equation}
    \sin^2 2 \theta_{23} = \dfrac{c^2}{1 + c^2} \,
\end{equation}
when $r=1$.
Thus the large mixing angle is
only realized when $c\gg 1$ (case (a)) where the 
large mixing is stable against
quantum corrections.
 For the Majorana phase, we can obtain $\cos \phi =-1$
from eq.(\ref{Majo1}) at $\chi=\pi/2$.
In this case,
the Majorana phase does not depend on $r$ and $c$.
Therefore the Majorana phase is 
independent of the energy-scale.
\vspace{24pt}

%
%

\section{Summary}

In this paper we study
 the energy-scale dependence of the MNS matrix
 with the physical Majorana phase
 in two generation neutrinos.
According to the LSND experiment\cite{LSND},
the scenario of two heavy degenerate neutrinos
 can be realistic and important,
 where the first generation effects are
 neglected in the energy-scale dependence of the MNS matrix.
In this case,
 the MNS matrix has two physical parameters, which are
 the mixing angle $\theta_{23}$ and the Majorana phase $\phi$.
The energy-scale dependence of these parameters
 are controlled by only one parameter $r$
 which is the ratio of the diagonal elements
 in the neutrino mass matrix \cite{HMOS1}.
This $r$ also governs
 the mass-squared difference scaled
 by the heaviest neutrino mass.
We can easily solve the renormalization group equation of
 this parameter, $r$.
Then the analytic solutions to the energy-scale dependence of
 the physical parameters can be obtained.
Especially, the energy-scale dependence of the Majorana phase
 is clarified by using these analytic solutions.
The instability of the Majorana phase causes in the same parameter
 region in which the mixing angle is unstable against quantum corrections.

%
%

\section*{Acknowledgment}
One of author NO is supported by the JSPS Research Fellowship
for Young Scientists, No.2996.

%
%

\end{document}